# A photonic-assisted method based on the MDA technique for the frequency estimation precision improvement

Gao guangyu[*][a], Liu naijin[*][a]

[a]Qian Xuesen Laboratory of Space Technology, China Academy of Space Technology, No.104 Youyi Street, Haidian District, Beijing, China
*gaoguangyu@qxslab.cn; liunaijin@qxslab.cn

**Abstract:** A novel photonics-assisted method based on presampling and MDA technique is proposed for significantly improving the frequency estimation precision without introducing other complex algorithms. This method is also compatible with existing FFT-based high-precision estimation algorithms. © 2019 The Author(s)
**OCIS codes:** 000.0000, 999.9999. [8-pt. type] For codes, see http://www.osapublishing.org/submit/ocis/

## 1. Introduction

Fast Fourier transform (FFT) is the most popular algorithm for spectrum analysis. However, suffering from the picket fence effect and spectrum leakage due to finite sampling time length, windowing and incoherence sampling, FFT technique usually has a coarse frequency resolution[1,2], and only locates the frequency at an approximately value with a maximum deviation up to half of the frequency resolution. Thus, a variety of methods have been developed to improve the frequency estimation precision in condition that not changing the sampling time length. To our knowledge, nearly all the efforts are focused on various frequency estimation algorithms in digital signal processing (DSP) at the back-end [3-7]. Although these algorithms work effectively more or less, preprocessing of signal at the front-end should be taken into account, for that this procedure would not only further enhance the estimation precision, but also is able to greatly reduce their computational cost. In this paper, we present a novel method based on presampling and multiorder deviation average technique (MDA) technique to implement the signal preprocessing procedure at the front-end. By applying photonic-assisted sampling and a simple FFT-based estimation algorithm based on averaging the measurement deviations in multiple Nyqusit zones, the proposed method can significantly improve the frequency estimation precision, needless to introduce other complex algorithms. Furthermore, this method as a representative preprocessing procedure is also compatible with other existing FFT-based frequency estimation algorithms.

## 2. Principle

The principal principle of the proposed method is based on presampling and MDA technique. For a single tone, its frequency $f_{in}$ can be expressed as $f_{in}=(m+\delta)f_{res}$, where $f_{res}=f_s/N$ is the FFT resolution with the sampling rate at $f_s$ and the sampling size at $N$. $m$ and $0\leq\delta<1$ are the integer part and the fractional part of the frequency index of $f_{in}$, denoting the position of $f_{in}$ in the spectrum. After sampling and FFT, the measurement value of $f_{in}$ is given by $f_{inM}=(m+[\delta])f_{res}$, where $[x]$ denotes rounding off to the nearest integer value. The measurement deviation of the original frequency is $\Delta^1 = f_{inM} - f_{in} = ([\delta] - \delta)f_{res}$. Due to $[\delta]=\text{int}(\delta)+[\text{rmod}(\delta)]$, where $\text{int}(x)$ and $\text{rmod}(x)$ denote the integer part and the fractional part of $x$, thus $\Delta^1$ can be simplified as $([\text{rmod}(\delta)] - \text{rmod}(\delta))f_{res}$. Apparently, the maximum of $\Delta^1$ is $\pm 0.5 f_{res}$. If the input signal is sampled by a presampling sequence with a repetition rate at $f_c$, the frequency of which will spread into the whole spectrum of the presampling sequence as shown in fig.1. Each Nyquist zone of the presampling sequence will contain a copy of the input signal. The signal after presampling can be filtered out by an anti-aliasing filter, and be sampled and quantitated by an electric ADC with a sampling rate at $f_s$.

If the original signal locates in the $2n+1$ order Nyquist zone, the frequency of its copy in the first Nyquist zone can be expressed as

$$f_{in}^n = f_{in} - nf_c = (m+\delta)f_{res} - n(\alpha+\varepsilon)f_{res} \qquad (1)$$

where $nf_c = n(\alpha+\varepsilon)f_{res}$, is the $n^{\text{th}}$ frequency line of the presampling sequence as shown in fig 1. $\alpha$ and $0\leq\varepsilon<1$ are the integer part and the fractional part of the frequency index of $f_c$. The measurement value of $f_{in}^n$ is given by

$$f_{inM}^n = \left[m+\delta-n(\alpha+\varepsilon)\right]f_{res} = f_{in} - nf_c + \left(\left[\text{rmod}(\delta-n\varepsilon)\right] - \text{rmod}(\delta-n\varepsilon)\right)f_{res} \qquad (2)$$

Equation (2) indicates that if the Nyquist zone where the frequency of the input signal locates in is known, it is easy to calculate the original frequency of the input signal based on equation (1) and (2), given as $f_{inM} = f_{inM}^n + nf_c$.

The measurement deviation of $f_{inM}$ is $\Delta^n=([\text{rmod}(\delta-n\varepsilon)]-\text{rmod}(\delta-n\varepsilon))f_{res}$. Apparently, $\Delta^n$ with the maximum at $0.5f_{res}$ depends on both $\delta$ and $\varepsilon$. Without loss of generality, $\Delta$ should be with weak dependence on $\varepsilon$. Since $\Delta$ is related to the ones from other different order Nyquist zones for they sharing the same $\varepsilon$ with different order number $n$, finding an appropriate $\varepsilon$ value and calculating the average value of the measurement values from multiple Nyquist zones should be a possible approach to reduce the dependence of $\Delta$ on $\varepsilon$ and improve the measurement precision.

The measurement average value $f^a_{inM}$ from $1^{th}$ to $N^{th}$ is $\sum_0^{N-1} f_{inM}/N = f_{in}+\Delta^a$, where $\Delta^a = \sum_0^{N-1}([\text{rmod}(\delta-n\varepsilon)]-\text{rmod}(\delta-n\varepsilon))f_{res}/N$, is the average value of $\Delta^n$, and $n \in [0, N-1]$ and $N \in \mathbb{N}$. The numerical results of the distributions of $\Delta^1/f_{res}$ and $\Delta^a/f_{res}$ versus $\delta$ are shown in figure 2, where $\varepsilon=0.1$ and $\varepsilon N=1$, revealing a significant precision improvement of $\Delta^a$ as compared with $\Delta^1$.

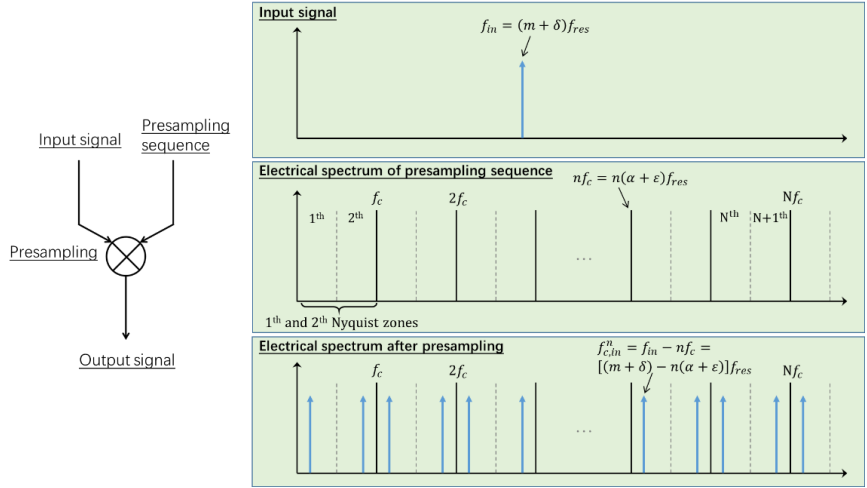

Figure 1 the principle diagram of the preprocessing procedure

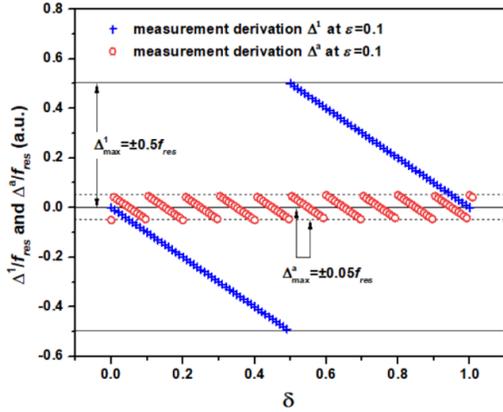

Figure 2 The distributions of $\Delta^1$ and $\Delta^a$ versus $\delta$

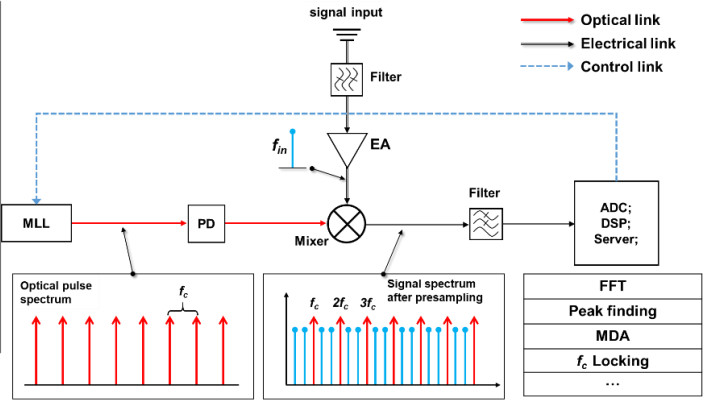

Figure 3 the schematic diagram of the proposed method. PD: Photoelectric detector; EA: electrical amplifier; MLL: mode-locked laser; MDA: multiorder deviation average

## 3. Results and Discussions

The proof-of-concept simulation based on the architecture as shown in Fig. 3 is implemented with Matlab to evaluate the feasiblity of the proposed method. The ultrashort optical pulse sequence generated from the mode-locked laser (MLL) is competent for the presampling process for which has a flat and ultra-wide spectrum distribution and a high-stability repetition rate. The optical pulses from MLL with a repetition rate $f_c$ at 100.02 MHz are directly converted by a high-speed and linear PD to its electrical pulse copy. The electrical pulse sequence presamples the input signal containing two single frequency tones with frequencies at 1.321 GHz and 3.774 GHz and SNR in spectral domain at 77dB on the electrical mixer, as shown in fig 3. Then, the signal are quantitated by the electric ADC with a sampling rate $f_s$ at 20 GSa/s, and transformed into spectral domain by FFT as shown in fig. 4, where the signal components in different Nyquist zones are clearly acquired, and several of them are used to calculate the fundamental frequencies. The sample size used for FFT is 1e5, giving a system resolution $f_{res}$ at 200 kHz and the frequency indexes of $f_c$ at $\alpha$=500 and $\varepsilon$=0.1, respectively. The integer frequency indexes of the signal components with frequency at $f_{inM}^n$ are found with the findpeaks module, and then one can calculate the frequency estimation value of the original signal. The measurement deviations $\Delta^n$ and the average value $\Delta^a$ are shown in fig 5. The distribution of $\Delta^n$ depends on the order $n$ and has a maximum at 100 kHz, while $\Delta^a$ as a constant at 10 kHz is independent on $n$, which is a 10 time improvement relative to $\Delta^n$. Actually, both $\Delta^n$ and $\Delta^a$ depend on a threshold value of SNR, as will be analyzed in the future work.

The simulation results preliminarily indicate that the proposed method is feasible in principle to improve the frequency estimation precise by a. In addition, this method is also compatible with other existing FFT-based high-precision frequency estimation algorithms, for example FFT-Quad Estimator. By using the FFT-Quad Estimator which executes a more precise peak finding simply though quadratic polynomial fitting after MDA calculation, the RMS of $\Delta^a$ decreases to about 120 Hz which is an improvement by a factor of over 800.

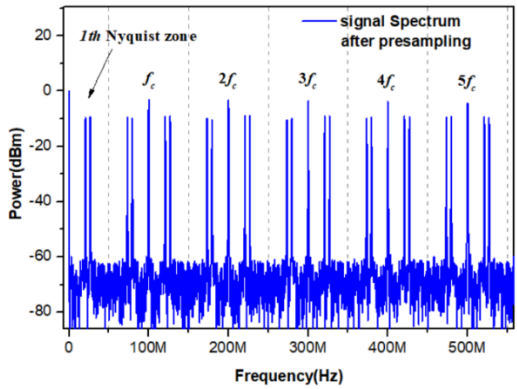
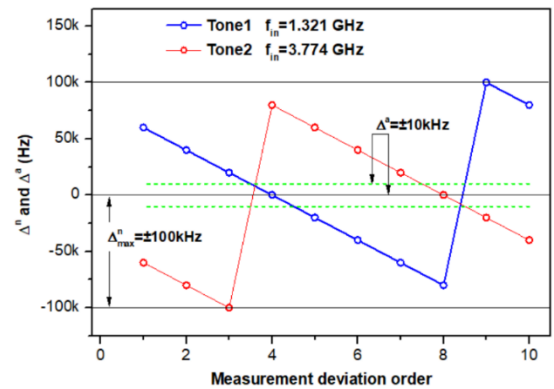

Figure 4 the signal spectrum of two tones after presampling

Figure 5 the measurement deviation $\Delta^n$ of different orders and their average value $\Delta^a$

## 4. Conlusion

In conclusion, a novel high-precision frequency estimation method mainly based on the photonic-assisted presampling and MDA technique is proposed in theory and is demonstrated in simulation. For multi-frequency tones with relatively high SNR, the proposed method can significantly improve the frequency estimation precision by 10 times, and even up to 800 times when other existing algorithm is employed after MDA calculation. This method has the potential for various application scenarios, such as Radar/LIDAR, spectrum sensing, vibration measurement and electronic reconnaissance.